# A fusion algorithm for joins based on collections in Odra (Object Database for Rapid Application development).

Mrs.Laika Satish[1], Dr.Sami Halawani[2]

[1]Lecturer, Faculty of Computing and IT, King Abdul Aziz University,
Rabigh, Saudi Arabia

[2]Dean, Faculty of Computing and IT, King Abdul Aziz University
Rabigh, Saudi Arabia

**Abstract**
In this paper we present the functionality of a currently under development database programming methodology called ODRA (Object Database for Rapid Application development) which works fully on the object oriented principles.
The database programming language is called SBQL (Stack based query language).
We discuss some concepts in ODRA for e.g. the working of ODRA, how ODRA runtime environment operates, the interoperability of ODRA with .net and java .A view of ODRA's working with web services and xml. Currently the stages under development in ODRA are query optimization. So we present the prior work that is done in ODRA related to Query optimization and we also present a new fusion algorithm of how ODRA can deal with joins based on collections like set, lists, and arrays for query optimization.

## 1. Introduction

Programming with object-oriented techniques is good since it works in concurrence with our experience in our real life, due to which it becomes simple and straightforward to visualize programs in a way in which objects interact with each other. The real objective of OO development is to create maintainable applications in the face of volatile requirements over the product life cycle in an independent platform.

An object conceals the specifics of the work accomplished but offers simple-to-use, standardized methods for performing particular operations on its data.

The key reason for the invention of object oriented methodology is to remove some flaws that the software engineers faced with the procedural approach .
To list a few important features of OOPS ,they are Scalable,Reusable and maintainable. Many who have build medium to large projects find that visualizing data types and processes as objects and object interactions makes describing systems and prototyping them much easier.

OOP matches very well the structures we create in this sort of visualization. RAD takes benefit of automated tools and techniques to restructure the process of information systems on the basis of OOPS. With lots of research on the comparison of approaches, techniques and tools, a proposal for a combined approach was given. This proposal is based upon the ideals of OO and that a
Combination could prove a solution to resolving these ideals. This proposal resulted in the invention of ODRA (Object Database for Rapid Application development) .It
is the next generation object-oriented environment for database management system based on SBA (Stack-Based Architecture). ODRA is currently under development at the Polish-Japanese Institute of Information Technology (Warsaw, Poland) since March 2004.

ODRA introduces a new universal declarative query and programming language SBQL (Stack-Based Query Language) [1,2,3,5,6], together with distributed, database-oriented and object-oriented execution environment. Such an approach provides functionality common to the variety of popular technologies (such as relational/object databases, several types of middleware, general purpose programming languages and their execution environments) in a single universal, easy to learn, interoperable and effective to use database application-programming environment. An object query is a query through which we can extract data from the object database. The result can be of type: objects, values or OID. In relational model, there are only two concepts: the relation and the attribute. The result of a query is always a new relation (with its attributes). In object model, there are more concepts: objects, values, OIDs, methods.

## 2. THE NEED FOR ODRA

For the next generation, database programmers there is a need for single, common, easy to learn, completely interoperable and effective to use database and application programming environment. In the areas of distributed and complex environments, programmers find it difficult to come up with the ideas for building applications that involves massive data processing.

Each time, the number of technologies, APIs, DBMSs, languages, tools, servers, etc., which the database programmers should learn and use, is extremely huge. This results in enormous software complexity, extensive costs and time of software manufacturing and permanently growing software maintenance overhead [2].

There has to be tools and a methodology for filters/wrappers to XML, web services, integrating diverse data sources and so on.

With the growth of non-classical database application, especially in the rapidly growing Internet context, the issue of bulk data processing in distributed and heterogeneous environments is becoming more and more important. Currently, increasing complexity and heterogeneity of such kinds of software has led to a situation where programmers are no more able to grasp every concept necessary to produce applications that could efficiently work in such environments. Therefore the research on new, simple, universal and homogeneous ideas of software development tools is currently very essential.

Most of the above is a need for an object oriented database query programming language.

All of these lead to the research of SBQL (Stack-Based Query Language), which is based on Stack, based approach, an object-oriented query and programming language. It is a query language with the full computational power of programming languages. SBQL alone makes it possible to create fully-fledged database and object oriented applications. It is a powerful query language extended to a programming language [2].

## 3. WORKING OF ODRA AND SBQL RUNTIME ENVIRONMENT

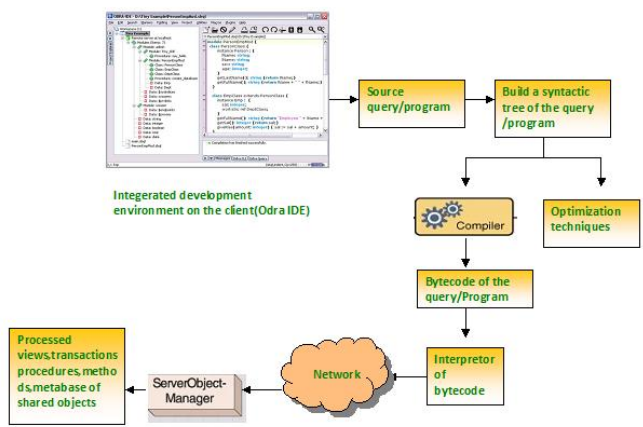

Fig.1 Working of ODRA (Object oriented database for Rapid application development)

**Step 1:** A source code of a query/program is created within the Integrated Development Environment, which includes an editor, a debugger, storage of source programs, storage of compiled programs, etc.

**Step2**: An abstract syntactic tree of the query or program is made with the help of a lexer and parser. The syntactic tree is checked for strong type checking, optimizations and the converted to bytecode which is ready to execute. This bytecode format is called Juliet.

**Step3**: Interpreter of bytecode. During runtime it takes instructions of a bytecode and triggers proper routines. To this end it uses two run-time stacks, ENVS (environment stack) and QRES (query result stack). The interpreter refers to volatile objects that are kept on a client and to any resources that are available on the server, in particular persistent (shared) objects. All the server resources are available through the object manager. [6]

The DBMS part controls the shared objects data store and provides all full functionality for transaction support, indexing, persistence, etc.

It is a main memory DBMS, in which persistence is based on modern operating systems' capabilities, such as memory mapped files. ODRA assumes the orthogonal persistence model [2]

Thus, The virtual machine (VM) in the SBQL execution environment uses the bytecode and provides services for hardware (virtual instruction set, virtual memory, etc.) and operating systems (loading, security, scheduling, etc.). After compilation, this bytecode can be run on every system for which ODRA has been ported. We plan that SBQL programs can also move from a busy computer to an idle computer during runtime.

## 4. ODRA INTEROPERABILITY FACILITIES (WITH .NET-NOBC AND JAVA-JOBC)

Use of java libraries via reflection: Java libraries can be accessed via reflection methods of java. The ODRA system offers many methods to access its resources from external applications with the help of SBQL.

**4.1 Java Object Base Connectivity (JOBC)**. This interface follows the style of the JDBC interface to relational databases. The differences concern input (SBQL rather than SQL) and output (serialized ODRA objects rather than serialized relational tables)

**4.2 .NET Object Base Connectivity for ODRA (NOBC)**

The NOBC (.NET Object Base Connectivity) is access client software to the ODRA system for the Microsoft .NET platform. It is similar to JOBC.
The software allows a .NET programmer to call SBQL queries and programs from .NET-based applications. As an inherent part the software allows the programmer to process the results of SBQL queries and programs on the side of .NET.

## 5. ODRA'S WORKING WITH XML AND WEB SERVICES

The ODRA system provides facilities for Web Services A mechanism called as ODRA proxy mechanism is applied to permit access to external data into the database
Remote services can be accessed through stubs in such a way that they are transparent to the SBQL environment that any mechanism can be applied to them, which makes remote and local resources act in the same way from the programmers view point. Dynamic Invocation without stubs can also be done. With XML, ODRA implements XML importer/exporter. Data can be derived from xml files as ODRA objects and can be processed with this powerful query language SBQL. XML files can be imported/exported without any loss of data. There are no limitations related to transformation of XML files.

## 6. SBQL4J

This is an extension to java language based on stack based architecture .It provides capabilities similar to Microsoft LINQ for Java language. It also allows to process Java objects with queries.

It's 100% compatible with current Java Virtual Machines, and can be safely used in any Java project (compatible with Java 6). [4]

## 7. DIFFERENT QUERY OPTIMIZATION METHODS IN SBQL.

Query processing is divided into query optimization and query execution. A query optimizer transforms a query written in a high-level query language into a series of processes that are implemented in the query execution engine.
The aim of query optimization is to find a query evaluation plan that increases performance, reduces I/O, and network time and effort, total resource usage, a combination of the above, or some other performance measure. Query optimization is a unique form of planning, utilizing techniques from artificial intelligence such as plan representation, dynamic programming, branch-and-bound algorithms. The motive of many query processing algorithms is to perform some kind of matching, i.e., bringing items that are "alike" together and performing some operation on them. [13]

There are different optimization techniques, which are available in ODRA like
*Query Optimization through Cached Queries for Object-Oriented Query Language SBQL.*
*Query Optimization by Result Caching in the Stack-Based Approach.*
*Query Optimization by Rewriting Compound Weakly Dependent Subqueries.*

## 8. A PROPOSED QUERY OPTIMIZATION ALGORITHM FOR SBQL

One of the differences between relational and object oriented databases (OODB) is that attributes in OODB can be of a collection type (e.g. sets, lists, arrays) as well as a simple type (e.g. integer, string).
Consider these join queries in SBQL
1. Retrieve student names with course names where they study:
   (Student join course). (fName, name)
2. Retrieve students who got more than 200 mks with their faculty names:
   ((Student where marks > 200) join (learns.faculty)).
    (fName, name)

We present an example *student* object that consists of attributes *fName*, *name,* and pointer links *learns*. Lets consider some attributes like address, which has sub attributes like *city, street* and house number. Each object, attribute, subobject, pointer, etc. has a unique internal identifier ($i_9$, $i_{10}$… $i_{22}$).

In object-oriented principles, A Collection consists of Arrays, Lists, and Set.

Arrays are one or many dimensional. Arrays are of h variable or fixed length, and in which duplicate values may exist. Sets are unordered collections but do not contain any duplicates. Lists are ordered collection, which may contain duplicate values. Our algorithm consists of joining the elements on the basis of hashing fundamentals.

For Example we take our university, which has many branches. Suppose we want to fetch the results for those students who study in the branches of Jeddah. Suppose there are 2 branches in Jeddah.

The join using hashing will first take all the branches in Jeddah and put them in a hash table in the main Memory, The student's table will be scanned and the hash value and the branch will be searched and compared with the values in the hash table in the main memory.

The fusion algorithm for joins based on collections:

Firstly the objects of the classes on which the join is based are divided on the basis of the first element for arrays and lists and if it's a set we take the minimum element from the set.

Then a merge process is done which consists of steps in which the array or the list or the set will be first sorted for each class in the join condition till we reach the end of classes involved in the join then we perform the hashing by hashing each object of the first class involved in the join into multiple hash tables.

For each object of the other classes involved in the join we will call a function.
A function which will take parameters element x and a hashing table y.
   Divide the elements of the collection into hash tables
   Hash and search for the element x in hash table y
   Boolean MatchResult;
   If a match is found Then
   X=x+1;y=y+1
      If finished searching for last part of the collection then
         MatchResult=true
         Return MatchResult
      End If
   Check for the existence of the hash table y
      If hash table y exist Then
         Recursively run the function
         Result = hashsearch (x, *y*)
      Else
         MatchResult=false
         Return MatchResult
      End If
   Else
      MatchResult=false
      Return MatchResult
   End If
      Return result

If this function returns true till we reach the collection in the end of the first class involved in the join and if match is found we put the identical pair into resulting output.

Graphical Evaluations

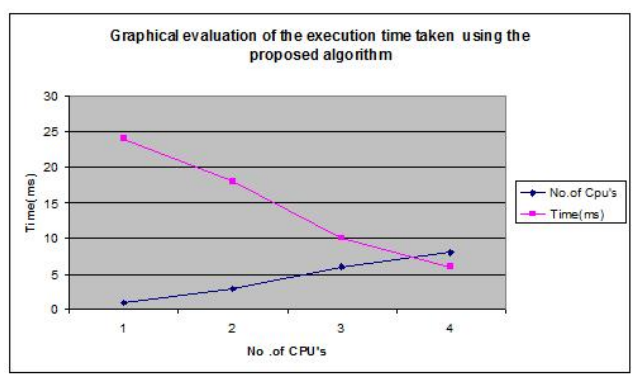

## 9. Conclusions

The main promising feature of this research into object oriented database query processing for new application areas is that the concept of a fixed number of parameters of the join, each doing the required data handling is flexible enough to meet the new challenges.

We can expect that much of the existing relational technology for query optimization and parallel execution will remain relevant, and that research into extensible optimization and parallelization will have a significant impact on future database application. ODRA should provide better modeling power, which will be widely accepted, but also competitive or better performance than previous systems.